\documentstyle[aasms4]{article}
\newcommand{\beq}{\begin{equation}}
\newcommand{\eeq}{\end{equation}}
\newcommand{\etal}{{\sl et~al.~}}

\def\fdg{\hbox{$.\!\!^\circ$}}
\def\arcmin{\hbox{$^\prime$}}

\begin{document}

\title{Astrometry with Hubble Space Telescope:
A Parallax of the Fundamental Distance Calibrator RR Lyrae\footnote{Based on 
observations made with
the NASA/ESA Hubble Space Telescope, obtained at the Space Telescope
Science Institute, which is operated by the
Association of Universities for Research in Astronomy, Inc., under NASA
contract NAS5-26555}}

\author{ G.\ Fritz Benedict\altaffilmark{1}, B. E.
McArthur\altaffilmark{1}, L.W.\
Fredrick\altaffilmark{12}, T. E. Harrison\altaffilmark{13}, J. Lee\altaffilmark{7}, C. L. Slesnick\altaffilmark{12}, J. Rhee\altaffilmark{12}, R. J. Patterson\altaffilmark{12},
E.\ Nelan\altaffilmark{5}, W.\ H.\ Jefferys\altaffilmark{6},  W.~van~Altena\altaffilmark{7}, P.~J.~Shelus\altaffilmark{1}, O. G. Franz\altaffilmark{2}, L.\ 
H. Wasserman\altaffilmark{2},
 P.D. Hemenway\altaffilmark{8}, R. L.
Duncombe\altaffilmark{9}, D. Story\altaffilmark{10}, A.\ L.\
Whipple\altaffilmark{10}, and A. J. Bradley\altaffilmark{11} }

\altaffiltext{1}{McDonald Observatory, University of Texas, Austin, TX 78712}
\altaffiltext{2}{Lowell Observatory, 1400 West Mars Hill Rd., Flagstaff, AZ 86001}
\altaffiltext{5}{Space Telescope Science Institute, 3700 San Martin Dr., Baltimore, MD 21218}
\altaffiltext{6}{ Department of Astronomy, University of Texas, Austin, TX 78712}
\altaffiltext{7}{ Department of Astronomy, Yale University, PO Box 208101, New Haven, CT 06520}
 \altaffiltext{8}{Department of Oceanography, University of Rhode Island, Kingston, RI 02881}
\altaffiltext{9}{Department of Aerospace Engineering, University of Texas, Austin, TX 78712}
\altaffiltext{10}{ Jackson and Tull, Aerospace Engineering Division
7375 Executive Place, Suite 200, Seabrook, Md.  20706}
\altaffiltext{11}{Spacecraft System Engineering Services, PO Box 91, Annapolis Junction, MD 20706}
\altaffiltext{12}{ Department of Astronomy, University of Virginia, PO Box 3818, Charlottesville, VA 22903}
\altaffiltext{13}{Department of Astronomy, New Mexico State University, Las Cruces, New Mexico 88003}



\begin{abstract}
We present an absolute parallax and relative proper motion for the fundamental distance scale calibrator, RR Lyr. We obtain these with astrometric data from
FGS 3, a white-light interferometer on {\it HST}. We find $\pi_{abs} = 3.82 \pm 0.2$ mas. Spectral classifications and VRIJHKT$_2$M and DDO51 photometry of the astrometric reference frame surrounding RR Lyr indicate that field extinction is low along this line of sight. We estimate $<$A$_V>=0.07\pm0.03$ for these reference stars. The extinction suffered by RR Lyr becomes one of the dominant contributors to the uncertainty in its absolute magnitude. Adopting the average field absorption, $<$A$_V>$=0.07 $\pm$ 0.03, we obtain M$_V^{RR} = 0.61 ^{-0.11}_{+0.10}$. This provides a distance modulus for the LMC, m-M=18.38 -- 18.53$^{-0.11}_{+0.10}$ with the average extinction-corrected magnitude of RR Lyr variables in the LMC, $<$V(RR)$>$, remaining a significant uncertainty. We compare this result to more than 80 other determinations of the distance modulus of the LMC.

\end{abstract}




%

\section{Introduction}

The various methods used to determine the distances to remote galaxies and ultimately the size, age, and shape of the Universe itself all depend on our knowledge of the distances to local objects. Among the most important of these are the RR Lyr variable stars. Considerable effort has gone into determining the absolute magnitudes, M$_V$, of these objects through statistical 
methods (e.g. \cite{Pop99}, \cite{Tsu98}, \cite{Pop98}, \cite{Fer98},
and \cite{Layden96}). For RR Lyr variables, this determination is 
complicated by dependence on metallicity, rendering the calibration uncertain (compare \cite{Fer98}, \cite{McN97}, \cite{Uda00a}, and \cite{Pop01}). Only recently has a relatively high-precision trigonometric parallax been available for RR Lyr from {\it HIPPARCOS} (RR Lyr = HIP 95497, \cite{Per97}). We have re-determined the parallax of RR Lyr  with FGS 3 on {\it Hubble Space Telescope} with higher precision. We hope to reduce zero point errors due to the spatially correlated errors in the {\it HIPPARCOS} catalog, discussed by  \cite{Nar99}. Additionally, our extensive investigation of the astrometric reference stars provides an independent estimation of the line of sight extinction to RR Lyr, a significant contributor to the uncertainty in its absolute magnitude, M$_V^{RR}$.

In this paper we describe the calibration allowing us to use a neutral density filter to relate astrometry of very bright targets to faint reference stars; present the results of extensive spectrophotometry of the astrometric reference stars, required to correct our relative parallax to absolute; briefly discuss data acquisition and analysis; and derive an absolute parallax for RR Lyr. Finally, we calculate an absolute magnitude for RR Lyr and apply it to derive a distance modulus for the Large Magellanic Cloud (LMC). We briefly review the present status of LMC distance moduli.

\cite{Bra91}
provide an overview of the
FGS 3 instrument and \cite{Ben99} describe the fringe tracking (POS) mode astrometric capabilities 
of FGS 3, along with the data acquisition and reduction strategies used in the present study. 
We time-tag our data with a modified Julian Date, $mJD = JD - 2444000.5$, and abbreviate millisecond of arc, mas, throughout.

\section{The Cross-Filter Calibration}

The filter wheel in each FGS contains a neutral density filter with a 1\% 
transmission (Nelan 2001). \nocite{Nel01} These filters, designated FND5, 
provide 5 magnitudes of attenuation. This reduction of signal is required to 
obtain astrometry for stars that are brighter than V = 8.5, for which the count 
rate for the FGS PMTs would exceed the electronics capacity (Bradley et al. 1991). No 
filter has perfectly plane-parallel faces, an effect called filter wedge. Filter 
wedge introduces a slight shift in position when comparing an observation with 
the standard astrometry filter, F583W, with the FND5 filter. We require 
this latter filter to perform astrometry on our primary science target, RR Lyr,  V 
$\sim$ 7.2. To obtain millisecond of arc astrometry requires knowledge of the 
filter wedge effect to that precision or better. \cite{Hem97} describe an early 
version of this calibration, but provide no explicit numbers, nor an 
estimate of the precision with which the calibration can be determined. Motivated by these difficulties, we obtained a second calibration in 1998. 

\subsection{Cross-Filter Calibration Observations}
Conceptually the calibration observations are simple. Observe in POS (fringe 
tracking) mode the same star with and without the FND5 filter and compare the 
positions. The shift so determined is then applied when comparing faint 
reference stars with bright science targets. The standard astrometry filter is F583W. As a consequence we will 
actually measure differential filter wedge, because F583W is also a filter with 
non-parallel faces. Note that each filter is a refractive element. Thus a star 
position will depend on the color of the star. This is an element of the lateral 
color effect discussed in \nocite{Ben99} Benedict et al. (1999). 

\begin{figure}
\epsscale{0.75}
\plotone{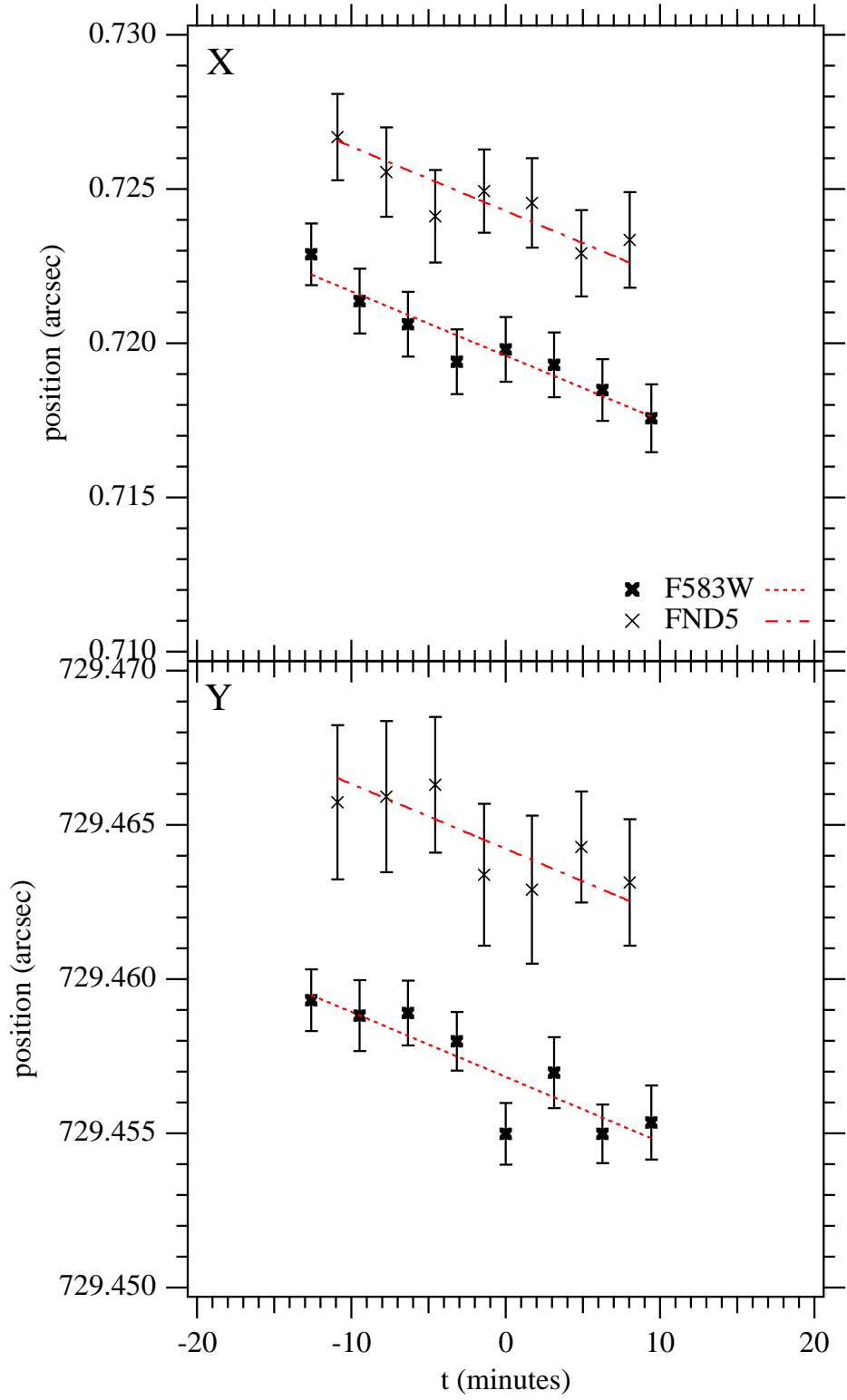}
\caption{Cross filter calibration observations in 1998. Target is Upgren 69 in NGC 188. The plots show the shift in position between F583W and FND5 and typical intra-orbit drift in FGS 3. } \label{fig-1}
\end{figure}

We carried out this calibration 
with two different stars at two different epochs. Our target in 1995 was HD 
41940 in M35. Our target in 1998 was Upgren 69 in the cluster NGC 188, a star 
utilized often as a template for fringe scanning with FGS 3
(Franz et al. 1998\nocite{Fra98}). We obtained eight to ten observations with F583W and 
seven observations with FND5 over 30 minutes on  14 March 1995 and 1 January 
1998. The 1998 measurements along each axis, X and Y, with 1$\sigma$ errors are 
plotted in Figure \ref{fig-1}. The errors associated with FND5 are larger because the signal from Upgren 69 is reduced by 99\%. These observations clearly show the 
offset due to differential filter wedge and a typical amount of intra-orbit positional drift.

\subsection{Cross-Filter Calibration Results}
We remove the effect of drift by fitting each set of measures with a line (see Figure~\ref{fig-1}) and 
adopt the offset between the lines at the midpoint of the sequence as the amount 
of differential filter wedge for X and Y, $\Delta$XFx and $\Delta$XFy. These corrections with associated 
error estimates, the cross-filter calibration, are collected in Table \ref{tbl-XF}, along with the actual attenuation in signal, $\Delta$m, due to the FND5 filter. Note 
that the differential filter wedge is different, comparing 1995 to 1998. This is 
due the lateral color effect discussed in Benedict et al. (1999). As can be seen 
in Table \ref{tbl-XF} the two calibration stars differ in color.

\section{Observations and Data Reduction}  \label{AstRefs}

Figure \ref{fig-2} shows the distribution in RA and Dec of the five reference stars and RR Lyr. Nine sets of data were acquired, spanning 2.09 years, for a total of 120 measurements  of  RR Lyr  and reference stars. Each data set required approximately 40 minutes of spacecraft time. The data were reduced and calibrated as detailed in \cite{Ben99} and \cite{mca01}. At each epoch we measured reference stars and the target, RR Lyr, multiple times, this to correct for intra-orbit drift of the type seen in the cross filter calibration data (Figure \ref{fig-1}).

\setcounter{footnote}{0}
\section{Spectrophotometric Absolute Parallaxes of the Astrometric Reference Stars}
Because the parallax determined for RR Lyr will be
measured with respect to reference frame stars which have their own
parallaxes, we must either apply a statistically derived correction from relative to absolute parallax (Van Altena, Lee \& Hofleit 1995, hereafter YPC95) or estimate the absolute parallaxes of the reference frame stars listed in Table \ref{tbl-POS}. In principle, the colors, spectral type, and luminosity class of a star can be used to estimate the absolute magnitude, M$_V$, and V-band absorption, A$_V$. The absolute parallax is then simply,
\beq
\pi_{abs} = 10^{-(V-M_V+5-A_V)/5}
\eeq

The luminosity class is generally more difficult to estimate than the spectral type (temperature class). However, the derived absolute magnitudes are critically dependent on the luminosity class. As a consequence we obtained additional photometry in an attempt to confirm the luminosity classes. Specifically, we employ the technique used by Majewski et al. (2000) to discriminate between giants and dwarfs for stars later than $\sim$ G5, an approach also discussed by  \cite{Pal94}.

\subsection{Photometry}
Our band passes for reference star photometry include: BVRI, JHK (from the second incremental release of 2MASS\footnote{ The Two Micron All Sky Survey
is a joint project of the University of Massachusetts and the Infrared Processing
and Analysis Center/California Institute of Technology }), and Washington/DDO filters M, 51, and T$_2$ (obtained at McDonald Observatory with the 0.8m Prime Focus Camera).  The 2MASS JHK have been transformed to the Bessell (1988) system using the transformations provided in \cite{Car01}. Tables \ref{tbl-VIS} and \ref{tbl-IR} list the visible, infrared, and Washington/DDO photometry for the RR Lyr  reference stars, RR-2 through RR-8.

\subsection{Spectroscopy}
The spectra from which we estimated spectral type and luminosity class come from WIYN\footnote{The 
WIYN Observatory is a joint facility of the University of
Wisconsin-Madison, Indiana University, Yale University, and the National
Optical Astronomy Observatories.} and New Mexico State University Apache 
Peak Observatory\footnote{ The
Apache Point Observatory 3.5 m telescope is owned and operated by
the Astrophysical Research Consortium.}. Classifications used a combination of template matching and line ratios. For this field we have two sets of spectral types and luminosity class for 4 out of 5 stars. Table \ref{tbl-SPP} lists the spectral types and luminosity classes for our reference stars. The differences between the WIYN and NMSU spectral types provide an estimate of $\sigma_{M_V}$. In those instances where the spectral types differ we adopt the classification closest to that suggested by a J-H vs H-K color-color diagram, the spectral type -- color mapping least affected by reddening. These colors are listed in Table~\ref{tbl-IR}.

The Washington/DDO photometry  provides a possible confirmation of the estimated luminosity class.
In Figure \ref{fig-3} we plot the Washington-DDO photometry along with a dividing line between dwarfs and giants (Paltoglou \& Bell 1994 \nocite{Pal94}). The boundary between giants and dwarfs is actually far 'fuzzier' than suggested by the solid line in Figure \ref{fig-3} and complicated by the photometric transition from dwarfs to giants through subgiants.  This soft boundary is readily apparent in Majewski et al. (2000) figure 14. Objects just above the heavy line are statistically more likely to be giants than objects just below the line. All but one of our reference stars lies on the dividing line to the left, where giant/dwarf discrimination is poorest. The remaining star, RR-5, moves closer to the other stars on the dividing line by correcting for the $<$A$_V>$=0.14 indicated in Table \ref{tbl-AV}. Except for this one measurement all the photometry is consistent with a dwarf classification for each reference star.

\subsection{Interstellar Extinction} \label{AV}
To determine interstellar extinction we first plot these stars on several color-color diagrams. A comparison of the relationships between spectral type and intrinsic color against those we measured provides an estimate of reddening. Figure \ref{fig-4} contains  V-R vs V-K and V-I vs V-K color-color diagrams and total galactic reddening vectors determined by \cite{Sch98} along this line of sight. Also plotted are mappings between spectral type and luminosity class V and III from \cite{Bes88} and \cite{Cox00} (hereafter AQ2000). again with maximum reddening vectors and the loci of luminosity class V and III stars. Figure~ \ref{fig-4}, along with the estimated spectral types, provides measures of the reddening for each reference star. 

\begin{figure}
\epsscale{1.0}
\plotone{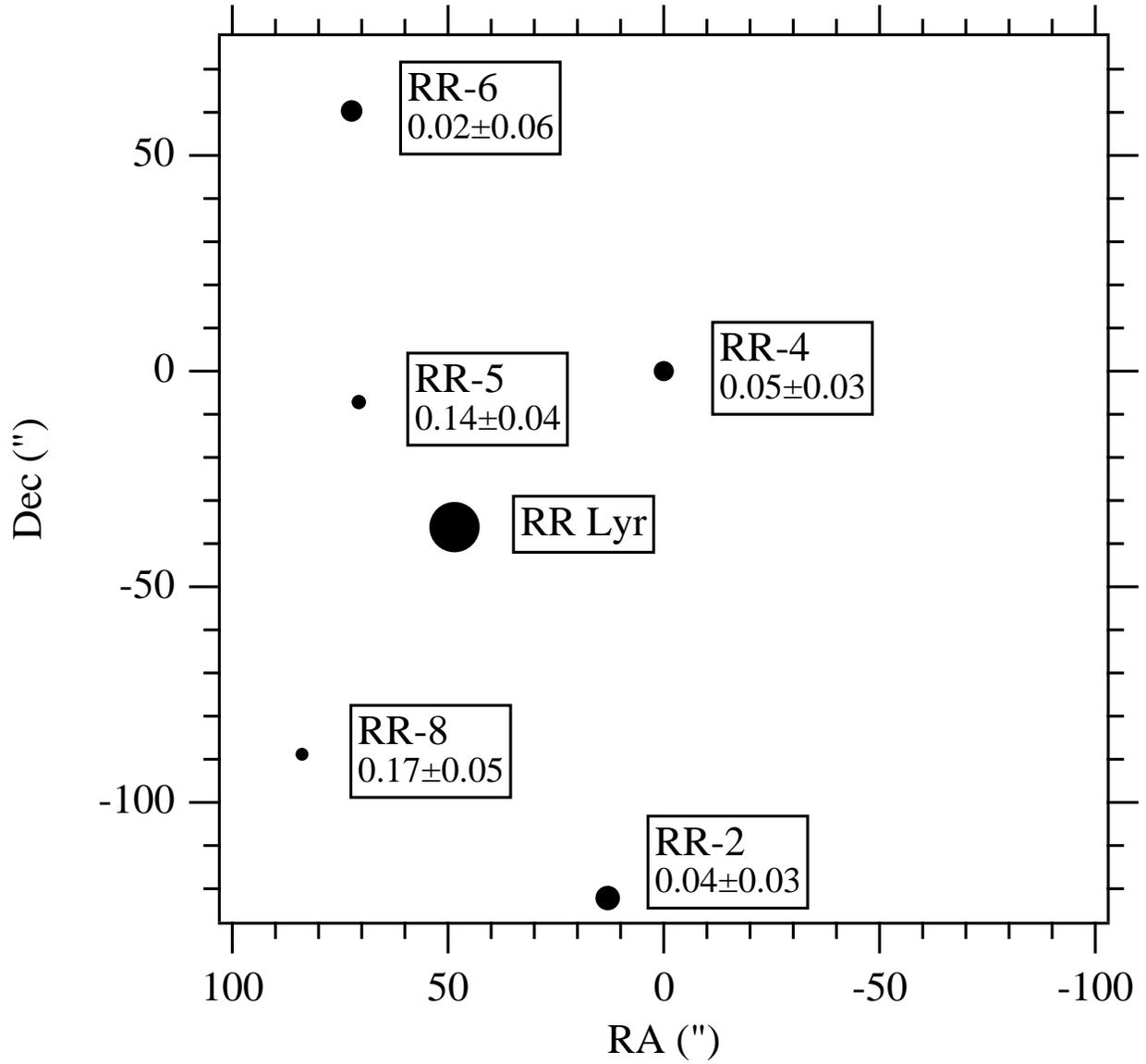}
\caption{RR Lyr and astrometric reference stars. Symbol size is indicative of V magnitude (Table~\ref{tbl-VIS}). The numbers within each identification box are the $<$A$_V>$ from Table \ref{tbl-AV}, Section \ref{AV}.}
\label{fig-2}
\end{figure}

\begin{figure}
\epsscale{1.0}
\plotone{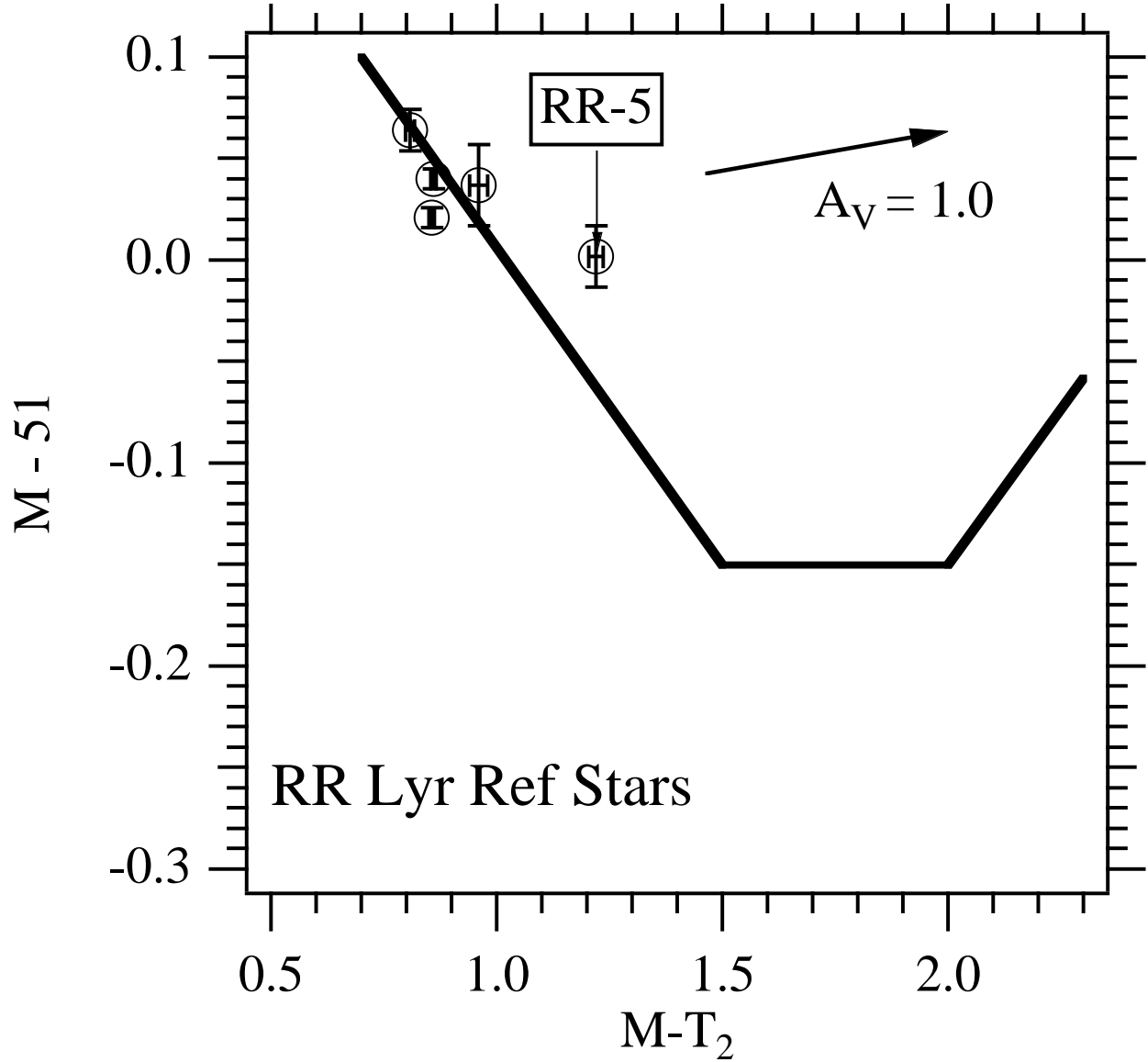}
\caption{M-DDO51 (M-51) vs M-T$_2$ color-color diagram. The solid line is the division between luminosity class V and luminosity class III stars. Giants are above the line, dwarfs below. The reddening vector is for A$_V$=1.0. De-reddening star RR-5 by the $<$A$_V> = 0.14$ value from Table~\ref{tbl-AV} would move it nearer to the dividing line between giants and dwarfs.}
\label{fig-3}
\end{figure}

\begin{figure}
\epsscale{0.65}
\plotone{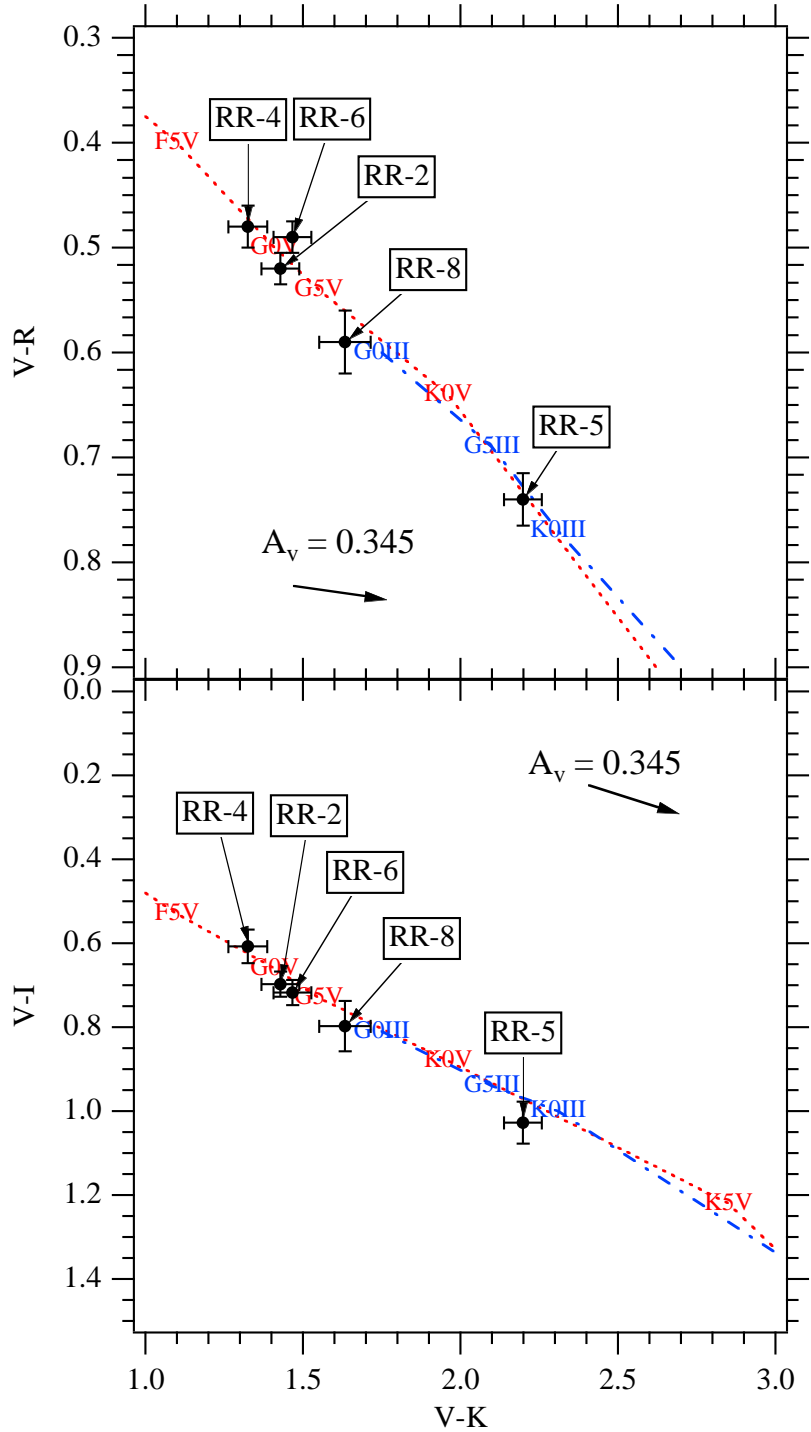}
\caption{V-R vs V-K and V-I vs V-K color-color diagrams. The dashed line is the locus of  dwarf (luminosity class V) stars of various spectral types; the dot-dashed line is for giants (luminosity class III). The reddening vector is the total galactic reddening determined by \cite{Sch98}.}
\label{fig-4}
\end{figure}

Assuming an R = 3.1 galactic reddening law (Savage \& Mathis 1977\nocite{Sav79}), we derive A$_V$ values by comparing the measured colors (Table~\ref{tbl-VIS}) with intrinsic V-R, V-I, and V-K colors from \cite{Bes88} and AQ2000. Specifically we estimate A$_V$ from three different ratios, each derived from the Savage \& Mathis (1977) reddening law: A$_V$/E(V-R) = 5.1; A$_V$/E(V-K) = 1.1; and A$_V$/E(V-I) = 2.4. These A$_V$ are collected in Table \ref{tbl-AV}. For the RR Lyr field colors and spectral types are consistent with a field-wide average $<$A$_V>$=0.07$\pm$0.03, far less than the maximum reddening, A$_V < 0.35$ determined by \cite{Sch98}. The spatial distribution of the average reddening for each star is shown in Figure \ref{fig-2}. We can only weakly assert that reddening is patchy, given the scatter in and uncertainty associated with these A$_V$ values. If we accept the variation in A$_V$ across the field,  a two dimensional linear interpolation in A$_V$ between the four nearest astrometric reference stars (Figure~\ref{fig-2}) yields A$_V$ = 0.11 $\pm$0.10 at the location of RR Lyr.

\subsection{Adopted Reference Frame Absolute Parallaxes}

We derive absolute parallaxes with M$_V$ values from AQ2000 and the $<$A$_V> $ derived from the photometry. Our parallax values are listed in Table \ref{tbl-SPP}. Individually, no reference star parallax is better determined than ${\sigma_{\pi}\over \pi}$ = 18\%. The average absolute parallax for the reference frame is $<\pi_{abs}>= 1.9$ mas.
As a check we compare this to the correction to absolute parallax discussed and presented
in the Yale Parallax Catalog (YPC95, Section 3.2, Fig. 2). Entering
YPC95, Fig. 2, with the RR Lyr galactic
latitude, l = 12\fdg5, and average magnitude for the
reference frame, $<$ Vref $>$= 13.75, we obtain a correction
to absolute of 1.1 mas. We will use the 1.9 mas correction derived from spectrophotometry, because the
use of spectrophotometric parallaxes offers a more direct way of
determining the reference star absolute parallaxes when such data are
available.

\section{Absolute Parallax of RR Lyr}
\subsection{The Astrometric Model}

With the positions measured by FGS 3 we determine the scale, rotation, and offset ``plate
constants" relative to an arbitrarily adopted constraint epoch (the so-called ``master plate") for
each observation set (the data acquired at each epoch). The mJD of each observation set is listed in Table~\ref{tbl-LOO}, along with a measured magnitude, a phase (based on P = 0.$^d$5668 (Kukarin \etal 1971\nocite{Kuk71}, rephased using the more recent photometry of \cite{Sch79}), and a B-V estimated by comparison with the UBV photometry of Hardie (1955)\nocite{Har55}. The RR Lyr reference frame contains 5 stars. We employ the six parameter model discussed in McArthur et al. (2001) for those observations. For the RR Lyr field all the reference stars are redder than the science target. Hence, we apply the corrections for lateral color discussed in Benedict et al. (1999). 

As for all our previous astrometric analyses, we employ GaussFit (\cite{Jef87}) to minimize $\chi^2$. The solved equations
of condition for RR Lyr are:
\beq
        x' = x + lcx(\it B-V) - \Delta XFx
\eeq
\beq
        y' = y + lcy(\it B-V) - \Delta XFy
\eeq
\beq
\xi = Ax' + By' + C + R_x (x'^2 + y'^2) - \mu_x \Delta t  - P_\alpha\pi_x
\eeq
\beq
\eta = -Bx' + Ay' + F + R_y(x'^2 + y'^2) - \mu_y \Delta t  - P_\delta\pi_y
\eeq
where $\it x$ and $\it y$ are the measured coordinates from {\it HST};
$\it lcx$ and $\it lcy$ are the
lateral color corrections from Benedict et al. 1999\nocite{Ben99}; and $\it B-V $ are
the  B-V  colors of each star, including the variable B-V of RR Lyr (Table~\ref{tbl-LOO}). $\Delta$XFx and $\Delta$XFy are the cross filter corrections in X and Y, applied only to the observations of RR Lyr. RR Lyr has a full range of  0.2 $<$ B-V $<$ 0.6. For this analysis we linearly interpolate between the 1995 and 1998 cross filter calibrations (Table~\ref{tbl-XF}) as a function of RR Lyr color.  A  and  B   
are scale and rotation plate constants, C and F are
offsets; $R_x$ and $R_y$ are radial terms;
$\mu_x$ and $\mu_y$ are proper motions; $\Delta$t is the epoch difference from the mean epoch;
$P_\alpha$ and $P_\delta$ are parallax factors;  and $\it \pi_x$ and $\it \pi_y$
 are  the parallaxes in x and y. We obtain the parallax factors from a JPL Earth orbit predictor (\cite{Sta90}), upgraded to version DE405. Orientation to the sky is obtained from ground-based astrometry 
(USNO-A2.0 catalog, Monet 1998\nocite{Mon98}) with uncertainties in the field orientation $\pm 0\fdg05$.
 
Solutions carried out constraining four reference stars to have no proper motion, allowing proper motion for the remaining reference star, indicate a statistically significant proper motion for reference star RR-5. Estimating the proper motion of that star imposed an 8\% decrease in number of degrees of freedom, but resulted in a 15\% decrease in $\chi^2$.

\subsection{Assessing Reference Frame Residuals}
\begin{figure}
\epsscale{0.6}
\plotone{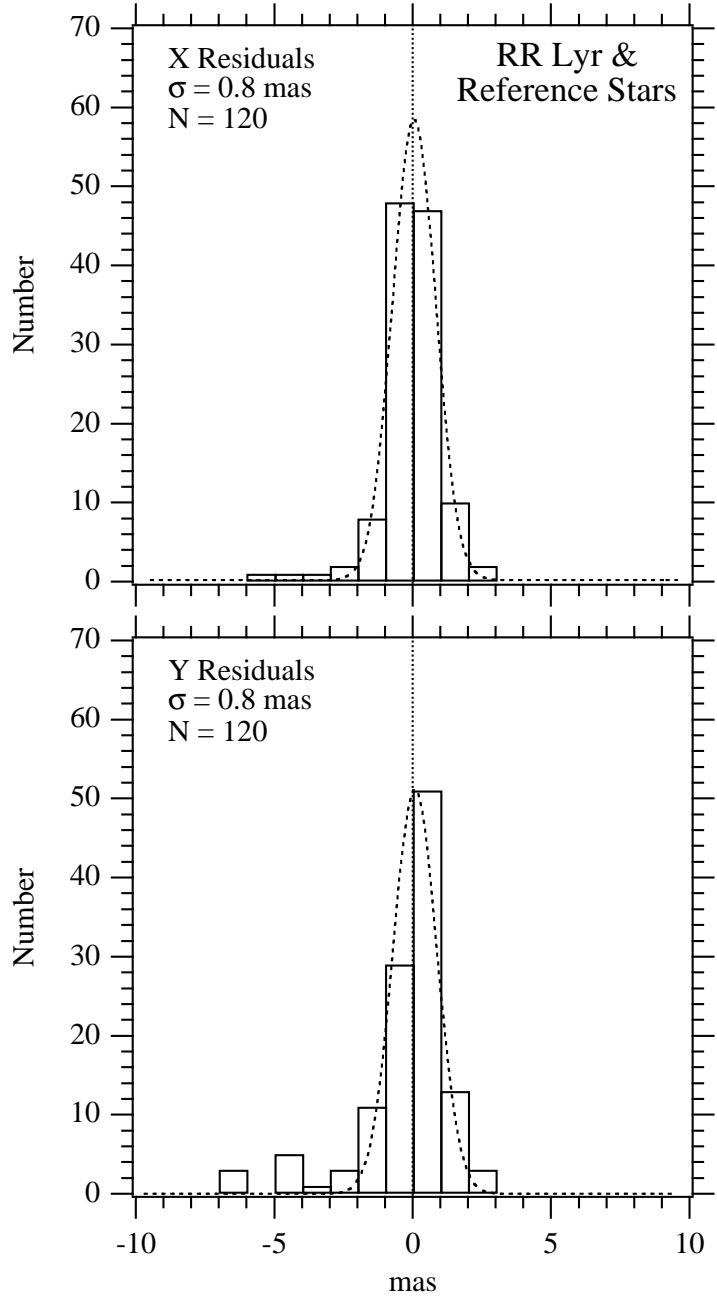}
\caption{Histograms of x and y residuals obtained from modeling RR Lyr and the astrometric reference stars with equations 4 and 5. Distributions are fit
with gaussians whose $\sigma$'s are noted in the plots.} \label{fig-5}
\end{figure}

The Optical Field Angle Distortion calibration (\cite{McA97}) reduces as-built {\it HST} telescope and FGS 3 distortions with magnitude $\sim1\arcsec$ to below 2 mas  over much of the FGS 3 field of regard. From histograms of the astrometric residuals (Figure~\ref{fig-5}) we conclude that we have obtained correction at the $\sim 1$ mas level in the
region available at all {\it HST} rolls (an inscribed circle centered on the pickle-shaped FGS field of regard). The resulting reference frame 'catalog' in $\xi$ and $\eta$ standard coordinates (Table \ref{tbl-POS}) was determined
with	$<\sigma_\xi>= 0.2$	 and	$<\sigma_\eta> = 0.3$ mas.

To determine if there might be unmodeled - but possibly correctable -  systematic effects at the 1 mas level, we plotted the RR Lyr reference frame X and Y residuals against a number of spacecraft, instrumental, and astronomical parameters. These included X, Y position within the pickle; radial distance from the pickle center; reference star V magnitude and B-V color; and epoch of observation.  We saw no obvious trends, other than an expected increase in positional uncertainty with reference star magnitude. 

\subsection{The Absolute Parallax of RR Lyr} \label{AbsPi}
In a quasi-Bayesian approach the calibration values were entered into the model as observations with associated errors.
The reference star spectrophotometric absolute parallaxes also were input as observations with associated errors, not as hardwired quantities known to infinite precision. 

\begin{figure}
\epsscale{1.0}
\plotone{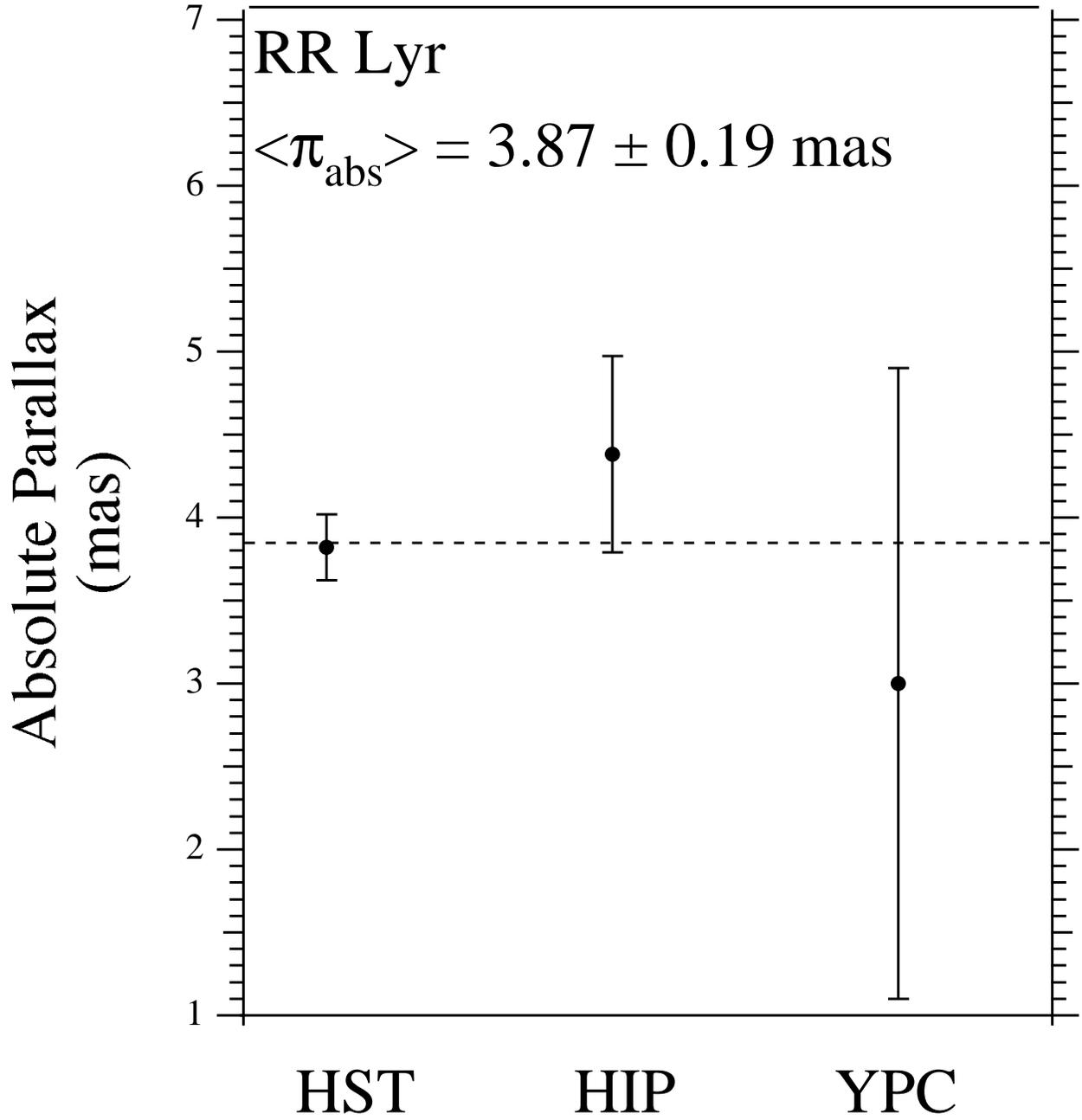}
\caption{Absolute parallax determinations for RR Lyr. We compare
{\it HST}, {\it HIPPARCOS}, and YPC95. The horizontal dashed line 
gives the weighted average absolute parallax, $<\pi_{abs}>$.} 
\label{fig-6}
\end{figure}

We obtain for RR Lyr an absolute parallax $\pi_{abs} = 3.82 \pm0.11$ mas, thus, ${\sigma_{\pi}\over \pi}$ = 3\%. This parallax differs by $\sim1\sigma_{\it HIP}$ and by $\sim4\sigma_{\it HST}$ from that measured by {\it HIPPARCOS}, $\pi_{abs} = 4.38 \pm0.59$ mas.
Comparing our various solutions with and without reference star proper motion we find some sensitivity in the resulting parallax, with a full range of variation of 0.2 mas. We feel this range represents a more likely error in our determination and adopt as the {\it HST} absolute parallax of RR Lyr, $\pi_{abs} = 3.82 \pm0.2$ mas (${\sigma_{\pi} \over \pi}$ = 4.6\%), an error one-third that of {\it HIPPARCOS}. Figure \ref{fig-6} compares the {\it HST}, {\it HIPPARCOS}, and the YPC95 (a weighted average of past ground-based results) determinations. The horizontal line is a weighted average of all three sources, $<\pi_{abs}> = 3.87 \pm0.19$ mas. Parallax and proper motion results from {\it HST}, {\it HIPPARCOS}, and YPC95 are collected in Table~\ref{tbl-SUM}.

\section{Discussion and Summary}
\subsection{{\it HST} Parallax Accuracy}
\begin{figure}
\epsscale{0.85}
\plotone{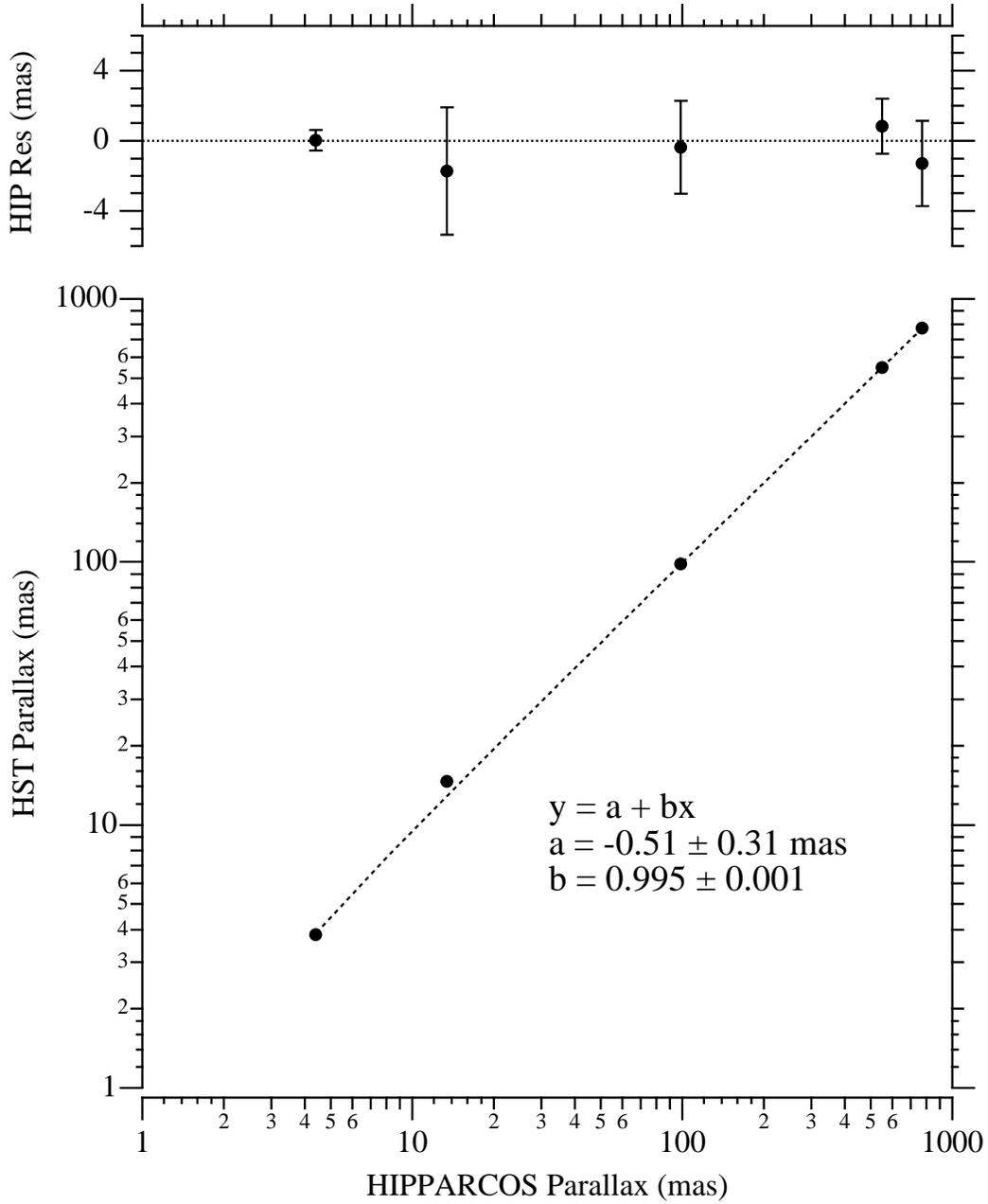}
\caption{Bottom: {\it HST} absolute parallax determinations compared
with {\it HIPPARCOS} for the five targets listed in Table~\ref{tbl-HH}. Top: The {\it HIPPARCOS} residuals to the error-weighted regression line. The error bars on the residuals are {\it HIPPARCOS} 1$\sigma$ errors. 
} \label{fig-7}
\end{figure}

Our parallax precision, an indication of our internal, random error, is often less than 0.5 mas. To assess our accuracy, or external error, we must compare our parallaxes with results from independent measurements. Following \cite{Gat98}, we plot all parallaxes obtained by the HST Astrometry Science Team with FGS 3 against those obtained by {\it HIPPARCOS}. These data are collected in Table \ref{tbl-HH} and shown in Figure \ref{fig-7}. We have not considered four Hyades stars whose parallaxes are considered preliminary (van Altena et al. 1997). The dashed line is a weighted regression that takes into account errors in both input data sets. The regression demonstrates the lack of scale and zero-point differences between {\it HIPPARCOS} and HST-FGS results. The rms {\it HIPPARCOS} residual to the regression line is 1.02 mas.

\subsection{The Lutz-Kelker Bias}

When using a trigonometric parallax to estimate the absolute
magnitude of a star, a correction should be made for the
Lutz-Kelker (LK) bias (\cite{Lut73}).
Because of the galactic latitude and distance of RR Lyr, 
and the scale height of the
stellar population of which it is a member,
we  use a uniform space density  for calculating
the LK bias.
An LK
algorithm modified by Hanson (H)(1979) that includes the power law
of the
parent population is used. A correction of -0.02 $\pm$ 0.01 mag is derived for
the
LKH bias for RR Lyr. The LKH bias is small because $\sigma_{\pi} \over \pi$= 4.6\% is small.

\subsection{The Absolute Magnitudes of RR Lyr}
Adopting for RR Lyr  
an intensity weighted average $<$V$>$= 7.76 (\cite{Fer98}) and  the absolute parallax weighted average from Section \ref{AbsPi} we determine that {\it in the absence of reddening} M$_V^{RR} = 0.68 ^{-0.10}_{+0.10}$, including the LKH correction and uncertainty. We derived (Section \ref{AV}) an average $<$ A$_V>$ = 0.07$\pm$0.03 from the astrometric reference stars that surround RR Lyr. Fernley et al. (1998) obtain for RR Lyr A$_V$ = 0.06$\pm$0.03 from a Log(P)-(V-K) relation. If there is no patchy extinction with angular scale less than 1\arcmin~it seems reasonable to conclude that RR Lyr, less distant than any reference star,  has A$_V \le $0.07 and, hence, M$_V^{RR} \ge 0.61 $. Including  this 0.03 magnitude uncertainty in $<$ A$_V>$ in quadrature, we obtain M$_V^{RR} = 0.61^{-0.11}_{+0.10}$. Alternatively, we could accept the A$_V$ variations seen in Figure~\ref{fig-2} as real and correct for a linearly interpolated A$_V$ = 0.11 $\pm$0.10, local to RR Lyr. Including that uncertainty in quadrature, we obtain M$_V^{RR} = 0.57 ^{-0.15}_{+0.14}$. 
Our range of values for M$_V^{RR}$ is remarkably close to that determined by \cite{Tsu98} from the {\it HIPPARCOS} parallax. We ascribe this similarity to differing LKH bias corrections and different A$_V$ corrections.

\cite{Bee00} cite an [Fe/H] - M$_V$ relation from Chaboyer (1999),
\beq
M_V^{RR} = 0.23([Fe/H]+1.6)+0.56 
\eeq
An [Fe/H]=-1.39 value for RR Lyr (also from Beers et al. 2000\nocite{Bee00}) implies M$_V^{RR}$=0.61, in agreement with either our M$_V^{RR}$ value derived from a variable A$_V$, or the M$_V^{RR}$ value derived from the average $<$ A$_V>$.

\subsection{The Distance Modulus of the LMC}
The distance to the Large Magellanic Cloud (LMC) is a critical link in
determining the scale of the universe. 
This distance uncertainty contributes  a substantial fraction of the uncertainty in
the Hubble Constant (\cite{Mould00}).
The HST Key Project on the Extragalactic Distance Scale (\cite{Freed01},
\cite{Mould00}) 
and the Type Ia Supernovae Calibration Team (\cite{Saha99}) have adopted
the distance modulus value m-M = 18.5.  Values from 18.1 to 18.8 are reported in the current
literature, with those less than 18.5 supporting the short distance scale and 
greater than 18.5, the long distance scale.
Comprehensive reviews of the methods can be found in
\nocite{Car00A}Carretta \etal (2000a), \cite{Gib99}, and \cite{Cole98}. A representative sample of
opinion about the "best method" can be found in \cite{Pac01}, 
Popowski (2001)\nocite{Pop01}, \cite{Uda00b}  and \cite{Gould00}.

Let us proceed with the constant $<$ A$_V>$ result, M$_V^{RR} = 
0.61^{-0.11}_{+0.10}$, because of our two results for absolute magnitude, it has smaller formal errors. We adopt (from \cite{Car00A}) $<$V$>$= 19.11 for the RR Lyrae 
variables in the bar of the LMC and their assumed $<$[Fe/H]$>$=-1.5. Correcting for the differential variation of M$_V$ with [Fe/H] (utilizing the slope from Chaboyer 1999), we compute $<$V$>$= 19.14, correcting from $<$[Fe/H]$>$= -1.50 to [Fe/H]=-1.39. We 
obtain an LMC distance modulus m-M = 18.53 $\pm$0.12. 
Carretta \etal (2000b)\nocite{Car00B} address possible luminosity differences between horizontal
branch stars in globular clusters and in the field and conclude that there
is little difference, unless the masses are very different.
There is another source of uncertainty in the LMC distance modulus, the 
measured apparent magnitude of the RR Lyr population in the LMC corrected 
for extinction local to the LMC.  For example, 
adopting (\cite{Uda99}) $<$V$>$= 18.94 and $<$[Fe/H]$>$=-1.6 for the 
RR Lyrae variables in the LMC, we obtain an LMC distance modulus 
m-M = 18.38 $\pm$0.12, once again correcting $<$V$>$ to [Fe/H]=-1.39. 

\begin{figure}
\epsscale{1.00}
\plotone{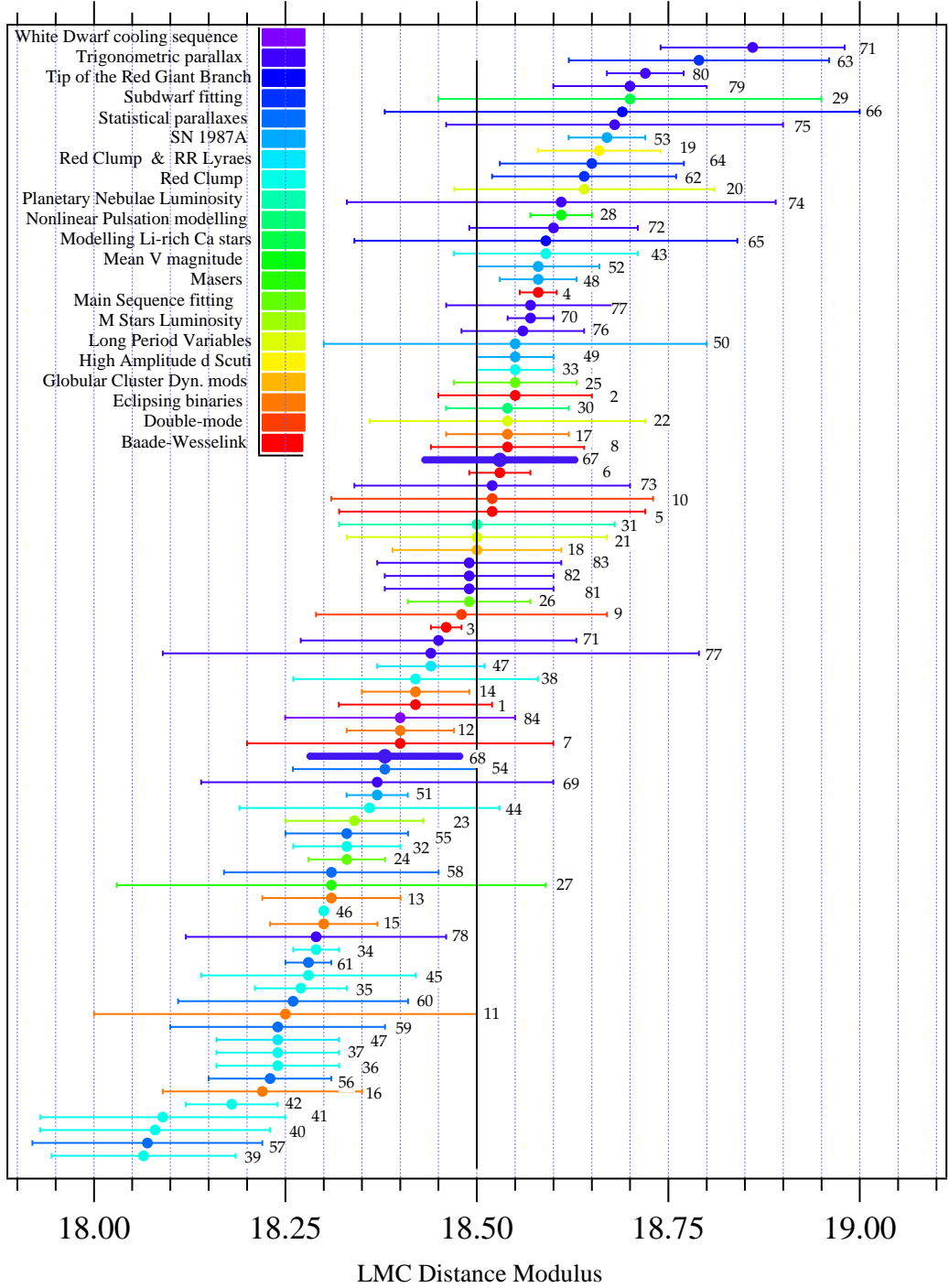}
\caption{Recent determinations of the distance modulus of the Large Magellanic 
Cloud, an expansion of the plot found in \cite{Gib99}.  Colors represent the various methods listed in column 2 of Table~\ref{tbl-LMC}, while the numbers refer to the individual investigations (column 1). Results from this paper are in bold. The thick vertical line denotes the distance modulus adopted by the HST Distance Scale Key Project (Freedman \etal 2001) and the Type Ia Supernovae Calibration Team (\cite{Saha99}.).
} \label{fig-8}
\end{figure}

These two estimates, which agree within their respective errors, are included in Figure~\ref{fig-8}, a summary of the current LMC distance modulus situation, displaying over 80 determinations,
based on 21 independent methods. These are listed
in Table~\ref{tbl-LMC}. For those cases with two associated uncertainties, the first is random error, the second systematic error. The weighted average of all these distance modulus determinations is $<$m-M$>$= 18.47$\pm$ 0.04, where the error is derived as the standard deviation of the mean with N=21.

\subsection{Summary}
{\it HST} astrometry yields an absolute trigonometric parallax for RR Lyr, $\pi_{abs} = 3.82 \pm0.2$ mas. A weighted average of {\it HST}, {\it HIPPARCOS}, and YPC95 absolute parallaxes is $<\pi_{abs}> = 3.87 \pm0.19$ mas. This high-precision result requires an extremely small Lutz-Kelker bias correction, -0.02$\pm$0.01 magnitude. Spectrophotometry of the astrometric reference stars local to RR Lyr suggest a low extinction, $<$ A$_V>$ = 0.07$\pm$0.03. The dominant error terms in the resulting absolute magnitude, M$_V^{RR} = 0.61 ^{-0.11}_{+0.10}$ ([Fe/H]=-1.39), are the parallax  and the uncertainty in the amount of extinction for RR Lyr itself. Depending on metallicity determinations and extinction corrections for RR Lyr variables in the LMC, we find a distance modulus range on the low end, m-M = [18.38 -- 18.53] $\pm$0.12, marginally supporting the `short scale' and an H$_0$ at the higher end of the present range.

\acknowledgments

Support for this work was provided by NASA through grants GTO NAG5-1603 from the Space Telescope 
Science Institute, which is operated
by the Association of Universities for Research in Astronomy, Inc., under
NASA contract NAS5-26555. These results are based partially on observations obtained with the
Apache Point Observatory 3.5 m telescope, which is owned and operated by
the Astrophysical Research Consortium, and the 
WIYN Observatory, a joint facility of the University of
Wisconsin-Madison, Indiana University, Yale University, and the National
Optical Astronomy Observatories. This publication makes use of data products from the Two Micron All Sky Survey,
which is a joint project of the University of Massachusetts and the Infrared Processing
and Analysis Center/California Institute of Technology, funded by the National
Aeronautics and Space Administration and the National Science Foundation. 
This research has made use of the SIMBAD database, operated at CDS, Strasbourg, France; the NASA/IPAC Extragalactic Database (NED) which is operated by the
            Jet Propulsion Laboratory, California Institute of Technology, under contract with the National Aeronautics
            and Space Administration;  and NASA's Astrophysics
Data System Abstract Service.
Thanks to Tom Barnes for helpful discussions and an early review of the text.
\clearpage


%

\clearpage
\begin{center}
\begin{deluxetable}{lllllccc}
\tablewidth{0in}
\tablecaption{Cross-filter Calibrations     \label{tbl-XF}}
\tablehead{\colhead{Year}&
\colhead{mJD} &
\colhead{star} &
\colhead{V} &
\colhead{B-V} &
\colhead{$\Delta$XFx ~~~(mas)} &
\colhead{$\Delta$XFy} &\colhead{$\Delta$m} }
\startdata
1995&49790.762&HD 41940& 8.14& 0.05 &  -3.5 $\pm$ 0.7& -6.9 $\pm$ 0.8 & 5.29\nl
1998&50814.813&Upgren 69&9.58& 0.50  &  -4.5  $\pm$      0.3& -7.2  $\pm$ 
0.5 & 5.21\nl
\enddata
\end{deluxetable}
\end{center}

\begin{deluxetable}{llllllll}
\tablewidth{0in}
\tablecaption{RR Lyrae  and Reference Star Data    \label{tbl-POS}}
\tablehead{\colhead{ID}&
\colhead{$\xi$ \tablenotemark{a}} &
\colhead{$\eta$ \tablenotemark{a}} &
\colhead{$\mu_x$ \tablenotemark{b}} &
\colhead{$\mu_y$ \tablenotemark{b}} }
\startdata
RR Lyr&
48.5557 $\pm$ 0.0002& -36.1748 $\pm$ 0.0003&-0.1061 $\pm$ 0.0002&-0.1973 $\pm$ 0.
0003\nl
RR-2  & 12.9877 $\pm$        0.0001& -122.0915 $\pm$ 0.0002\nl
RR-4\tablenotemark{c} &  0.0000 $\pm$  0.0002&  0.0000 $\pm$   0.0002\nl
RR-5& 70.6893 $\pm$ 0.0002& -7.2290 $\pm$ 0.0003&0.0045 $\pm$ 0.0003&-0.0123 $\pm$ 0.0003
\nl
RR-6  & 72.4101 $\pm$ 0.0002 & 60.4006 $\pm$  0.0004 \nl
RR-8  & 83.8194 $\pm$ 0.0002 & -88.7945 $\pm$ 0.0002 \nl
\enddata
\tablenotetext{a}{$\xi$ and $\eta$ are relative positions in arcseconds
}
\tablenotetext{b}{$\mu_x$ and $\mu_y$ are relative motions in arcsec
yr$^{-1}$ }
\tablenotetext{c}{RA = 19 25 23.56 Dec = 42 47 40.7, J2000, epoch = mJD
50201.05
711}
\end{deluxetable}

\begin{deluxetable}{llllllll}
\tablewidth{0in}
\tablecaption{Astrometric Reference Stars Visible Photometry
\label{tbl-VIS}}
\tablehead{\colhead{RR-ref}&
\colhead{V} &
\colhead{V-R} &
\colhead{V-I} &
\colhead{V-K} }
\startdata
2&  12.68  $\pm$ 0.02 &0.51 $\pm$ 0.03 & 0.68 $\pm$ 0.03& 1.429 $\pm$ 0.06\nl
4&  13.47 $\pm$   0.02&  0.48 $\pm$   0.04 & 0.61 $\pm$   0.04 & 1.325  $\pm$ 
0.06\nl
5&  14.50 $\pm$   0.02&  0.67 $\pm$   0.05 & 0.94 $\pm$   0.05 & 2.199 $\pm$ 
0.06\nl
6&  13.15 $\pm$   0.02&  0.49 $\pm$   0.03 & 0.70 $\pm$  0.03  & 1.467 $\pm$ 
0.06\nl
8&  14.94 $\pm$  0.02 &  0.56 $\pm$   0.06 & 0.76 $\pm$   0.06 & 1.633  $\pm$ 
0.08\nl
\enddata
\end{deluxetable}

\begin{deluxetable}{llllllll}
\tablewidth{0in}
\tablecaption{Astrometric Reference Stars Near-IR and Washington-DDO
Photometry
  \label{tbl-IR}}
\tablehead{\colhead{RR-ref}&
\colhead{K} &
\colhead{J-H} &
\colhead{H-K} &
\colhead{M-T$_2$} &
\colhead{M-51} }
\startdata
2&11.21 $\pm$ 0.03&0.31 $\pm$ 0.02 & 0.07 $\pm$ 0.02 & 0.86 $\pm$ 0.01
&0.02 $\pm$ 0.01\nl
4&12.11 $\pm$   0.03&0.29 $\pm$ 0.02&0.06 $\pm$ 0.02&0.81 $\pm$ 0.01
&0.06 $\pm$ 0.01\nl
5&12.26 $\pm$   0.03&0.53 $\pm$ 0.02&0.09 $\pm$ 0.02&1.22 $\pm$ 0.02
&0.00 $\pm$ 0.02\nl
6&11.64 $\pm$   0.03&0.33 $\pm$ 0.02&0.07 $\pm$ 0.02&0.86 $\pm$ 0.01
&0.04 $\pm$ 0.01\nl
8&13.27 $\pm$   0.04&0.37 $\pm$ 0.02&0.04 $\pm$ 0.02&0.96 $\pm$ 0.02
&0.04 $\pm$ 0.02\nl
\enddata
\end{deluxetable}

\begin{center}
\begin{deluxetable}{llllllll}
\tablewidth{0in}
\tablecaption{Reference Star A$_V$ from Spectrophotometry  \label{tbl-AV}}
\tablehead{  \colhead{RR-ref}&
\colhead{A$_V$(V-I)}&   \colhead{A$_V$(V-R)}&  \colhead{A$_V$(V-K)} &
\colhead{$<A$$_V$$>$}}
\startdata
2&0.07&0.05&-0.01&0.04 $\pm$ 0.03\nl
4&0.00&0.10&0.04&0.05 $\pm$ 0.03\nl
5&0.19&0.15&0.06&0.14 $\pm$ 0.04\nl
6&0.12&-0.10&0.03&0.02 $\pm$ 0.06\nl
8&0.17&0.26&0.09&0.17 $\pm$ 0.05\nl
\nl
$<$A$_V$$>$&0.09&0.08&0.04&0.07 $\pm$ 0.03\nl
\enddata
\end{deluxetable}
\end{center}

\begin{deluxetable}{llllllllll}
\tablewidth{0in}
\tablecaption{Astrometric Reference Star Spectral Classifications and
Spectrophotometric Parallaxes \label{tbl-SPP}}
\tablehead{\colhead{RR-ref}& \colhead{WIYN}&
\colhead{NMSU}&\colhead{adopted} &
\colhead{V} & \colhead{M$_V$} & \colhead{A$_V$} &
\colhead{$\pi_{abs}$} & \colhead{$\sigma_{\pi}\over\pi$(\%)} }
\startdata
2& G1 V& G3 V& G1 V&12.68&4.6 $\pm$ 0.4&0.07&2.5 $\pm$  0.5& 20 \nl
4& G0 V& F8 V& F8 V&13.47&4.0 $\pm$ 0.4&0.07& 1.3 $\pm$  0.2& 15 \nl
5& & K1 V& K1 V&14.50&6.1 $\pm$ 0.4&0.07& 2.2 $\pm$  0.4& 18 \nl
6& G1 V&G0 V& G1 V&13.15&4.6 $\pm$ 0.4&0.07& 2.0 $\pm$  0.2& 10 \nl
8& G2 V&G5 V& G5 V&14.94&5.1 $\pm$ 0.4&0.07& 1.1 $\pm$  0.2& 18 \nl
\enddata
\end{deluxetable}

\begin{deluxetable}{lllll}
\tablewidth{4in}
\tablecaption{RR Lyr Log of Observations\label{tbl-LOO}}
\tablehead{\colhead{Set}&
\colhead{mJD} &
\colhead{phase \tablenotemark{a}} &
\colhead{V \tablenotemark{b}} &
\colhead{B-V \tablenotemark{c}} 
}
\startdata
1&49984.76525&0.36&7.99&0.3\\
2&50047.5617&0.24&7.70&0.3\\
3&50173.4915&0.34&7.96&0.3\\
4&50201.57787&0.65&8.19&0.44\\
5&50229.3208&0.62&8.24&0.44\\
6&50563.39767&0.29&7.86&0.3\\
7&50568.10532&0.46&8.08&0.4\\
8&50745.70698&0.65&8.26&0.44\\
9&50749.74358&0.16&7.39&0.2\\
\enddata
\tablenotetext{a}{Phase based on P = 0.$^d$5668 (\cite{Kuk71})}
\tablenotetext{b}{From FGS photometry. See \cite{Ben98} for transformation details. }
\tablenotetext{c}{Estimated from phase using UBV photometry from Hardie (1955)\nocite{Har55}.}
\end{deluxetable}

\begin{center}
\begin{deluxetable}{ll}
\tablecaption{RR Lyr Parallax and Proper Motion \label{tbl-SUM}}
\tablewidth{0in}
\tablehead{\colhead{Parameter} &  \colhead{ Value }}
\startdata
{\it HST} study duration  &2.09 y\nl
number of observation sets    &   10 \nl
ref. stars $ <V> $ &  $13.75 $  \nl
ref. stars $ <B-V> $ &  $0.71 $ \nl
\nl
{\it HST} Absolute Parallax   & 3.82  $\pm$  0.20   mas\nl
{\it HIPPARCOS} Absolute Parallax &4.38  $\pm$ 0.59 mas\nl
YPC95 Absolute Parallax & 3.0  $\pm$  1.9  mas\nl
Weighted average Absolute Parallax & 3.87 $\pm$  0.19 mas \nl
\nl
{\it HST} Proper Motion  &224.0  $\pm$  0.5 mas y$^{-1}$ \nl
 \indent in pos. angle & 208\fdg3  $\pm$ 0\fdg 5 \nl
{\it HIPPARCOS} Proper Motion  &224.2  $\pm$  1.4 mas y$^{-1}$ \nl
 \indent in pos. angle & 209\fdg3  $\pm$ 1\fdg3 \nl
YPC95 Proper Motion  &207.4  mas y$^{-1}$ \nl
 \indent in pos. angle & 210\fdg7  \nl
\enddata
\end{deluxetable}
\end{center}

\begin{deluxetable}{llllllllll}
\tablewidth{0in}
\tablecaption{{\it HST} and {\it HIPPARCOS} Absolute Parallaxes   \label{tbl-HH}}
\tablehead{\colhead{ID}& \colhead{{\it HST}}& \colhead{{\it HIP}} &\colhead{{\it HST} Reference}}
\startdata
Prox Cen  &     769.7  $\pm$   0.3   & 772.33 $\pm$   2.42 & Benedict et al. 1999\nl
Barnard's Star& 545.5  $\pm$   0.3   & 549.3  $\pm$   1.58 & Benedict et al. 1999\nl
Feige 24      & 14.6   $\pm$   0.4   & 13.44  $\pm$   3.62 & Benedict et al. 2000\nl
Gl 748 AB     & 98.0   $\pm$     0.4   & 98.56  $\pm$   2.66 & Benedict et al. 2001\nl
RR Lyr &3.82   $\pm$   0.20 &   4.38  $\pm$     0.59 & this paper\nl
\enddata
\end{deluxetable}

\begin{center}        
\begin{deluxetable}{llllllllll}        
\tablewidth{0in}        
\tablecaption{Recent Distance Moduli to the LMC      \label{tbl-LMC}}        
\tablehead{\colhead{No.} & \colhead{Method}&  \colhead{Object} &  \colhead{Author}&  \colhead{m-M} }    
\startdata        
1 & Baade-Wesselink & Cepheids   & Gieren et al. 2000  & 18.42 $\pm$   0.10\nl 
2 &  &  Cepheids  & Carretta et al.2000 & 18.55 $\pm$   0.10\nl 
3 &  &Cepheid  &  Gieren et al. 1998& 18.46 $\pm$   0.02\nl   
4 &  &Cepheid  & Di Benedetto   1997 &18.58 $\pm$   0.024\nl  
5 &  & RR Lyraes  & Carretta et al. 2000  &  18.52   $\pm$    0.20 \nl 
6 &  &RR Lyraes  & Feast 1997 & 18.53 $\pm$   0.04 \nl   
7 &  &RR Lyraes  & Cacciari et al. 1990  &  18.40  $\pm$    0.20  \nl
8 &  & RR Lyraes  & McNamara 1997   &  18.54  $\pm$    0.10\nl 
9 & Double-mode  &  RR Lyraes  & Alcock  1997    &     18.48 $\pm$  0.19 \nl 
10 &  &  RR Lyraes  & Kovacs        2000   &     18.52 $\pm$  0.21 \nl 
11 & Eclipsing binaries  &  EROS 1044  &  Maloney et al. 2001  &  18.25 $\pm$  0.25 \nl 
12 &  &  HV 2274 &  Nelson 2000  & 18.40  $\pm$  0.07 \nl 
13 &  &  HV 982 &  Fitzpatrick 2000  & 18.31  $\pm$  0.09 \nl 
14 &  & HV 2274  &  Guinan et al. 1998a   & 18.42  $\pm$  0.07 \nl 
15 &  & HV 2274  &  Guinan et al. 1998b   & 18.30  $\pm$  0.07 \nl 
16 &  &  HV 2274 &  Udalski 1998a  & 18.22     $\pm$  0.13 \nl 
17 &  & HV2274 &  Guinan et al. 1997   & 18.54  $\pm$  0.08 \nl 
18 & Globular Cluster Dyn. mods &  &  Chaboyer et al. 1998 &  18.50  $\pm$  0.11\nl 
19 & High Amplitude $\delta$ Scuti& $\delta$ Scuti &  McNamara 2001  &  18.66  $\pm$  0.08\nl 
20 & Long Period Variables& Mira &Whitelock \& Feast 2000 & 18.64 $\pm$ 0.17 \nl
21 & &Ca Stars & Bergeat  et al. 1998  &  18.50 $\pm$ 0.17 \nl
22 &  & Mira  &  Van Leeuwen et al.  1997 & 18.54   $\pm$  0.18 \nl
23 & M Stars Luminosity  &  &  Schmidt-Kaler \& Oestreicher 1998  &  18.34 $\pm$ 0.09 \nl
24 & Main Sequence fitting    &  NGC 1866 & Walker et al. 2001 & 18.33 $\pm$  0.05    \nl
25 & &Cepheids  &  Carretta et al. 2000 &18.55   $\pm$  0.04 $\pm$ 0.04\nl  
26 &  & Cepheids & Laney \& Stobie  1994 & 18.49   $\pm$  0.04 $\pm$ 0.04 \nl
27 & Masers  & NGC 4258   &  Newman et al. 2001  &  18.31  $\pm$  0.11 $\pm$  0.17 \nl
28 & Mean V magnitude  &  LMC RR Lyraess & McNamara 2001  &  18.61  $\pm$    0.04  \nl
29 & Modelling Li-rich Ca stars&  &  Ventura 1999  &  18.70 $\pm$  0.25    \nl
30 & Nonlinear Pulsation modelling &  Cepheids   &  Wood  1998  & 18.54   $\pm$ 0.08 \nl
31 & Planetary Nebulae Luminosity  &  M31   &  Walker  1999  & 18.50   $\pm$ 0.18 \nl
32 & Red Clump  &    &  Popowski  2001  &  18.33 $\pm$ 0.07 \tablenotemark{a} \nl
33 &  &  &  Girardi \& Salaris  2001  &  18.55 $\pm$ 0.05  \nl
34&   &  &  Sakai 2000&18.29 $\pm$  0.03 \nl
35 &  &  &  Popowski  2000  &  18.27 $\pm$ 0.06 \tablenotemark{b} \nl
36 &  &  &  Stanek et al. 2000  & 18.24  $\pm$  0.08 \nl
37 &  &  &  Udalski  2000a  & 18.24  $\pm$  0.08 \nl
38 &  &  &  Twarog et al. 1999  & 18.42  $\pm$  0.16 \nl
39 &  &  &  Stanek et al. 1998  &  18.065 $\pm$ 0.12 \nl
40 &  &  &  Udalski et al.1998b  & 18.08  $\pm$  0.15 \nl
41 &  &  &  Udalski 1998c  & 18.09  $\pm$  0.16 \nl
42 &  &  &  Udalski 1998d  & 18.18  $\pm$  0.06 \nl
43 &  &  &  Romaniello et al. 2000  & 18.59  $\pm$  0.04 $\pm$ 0.08 \nl
44 &  &  &  Cole 1998  &  18.36  $\pm$  0.17 \nl
45 &  &  &  Girardi et al. 1998 & 18.28  $\pm$  0.14\nl  
46 &  &   &  Beaulieu \& Sackett 1998  &  18.3 \nl
47 & Red Clump \& RR Lyraes  &    &  Popowski 2001  & 18.24  $\pm$  0.08 \nl
47 &  &  &  & to \nl
47 &  &  &  & 18.44  $\pm$  0.07 \nl
48 & SN 1987A &    & Carretta et al. 2000  & 18.58 $\pm$  0.05 \nl
49 &  &  & Romaniello et al. 2000   & 18.55 $\pm$  0.05   \nl
50 &  &  & Walker  1999   & 18.55 $\pm$  0.07 $\pm$  0.16   \nl
51 &  &  &  Gould \& Uza 1998  &      18.37 $\pm$  0.04   \nl
52 &  &  & Panagia et al. 1998   & 18.58 $\pm$  0.08   \nl
53 &  &  & Lundqvist \& Sonneborn 1998 &  18.67 $\pm$  0.05    \nl
54 & Statistical parallaxes & RR Lyraes   & Carretta   et al. 2000 & 18.38 $\pm$  0.12 \nl 
55 &  & RR Lyraes  &  Popowski \& Gould 1999  &  18.33 $\pm$ 0.08 \tablenotemark{a} \nl 
56 &  & RR Lyraes &  Popowski \& Gould 1999  &  18.23 $\pm$ 0.08 \tablenotemark{b} \nl 
57 &  & RR Lyraes &  Popowski \& Gould 1998  &  18.07 $\pm$ 0.15 \tablenotemark{a} \nl 
58 &  & RR Lyraes &  Popowski \& Gould 1998  &  18.31 $\pm$ 0.14 \tablenotemark{b} \nl 
59 &  & RR Lyraes  &  Gould \& Popowski 1998  &  18.24 $\pm$  0.14 \nl 
60 &  & RR Lyraes  & Fernley et al.  1998  &  18.26 $\pm$ 0.15\nl 
61 &  & RR Lyraes  &  Layden et. al 1996  &  18.28 $\pm$ 0.03 \nl 
62 & Subdwarf fitting  &   &  Carreta et al. 2000  &      18.64 $\pm$  0.12    \nl 
63 &  &   &  Reid 1998  &   18.79 $\pm$  0.17    \nl 
64 &  &   &  Reid 1997  &   18.65 $\pm$  0.12    \nl 
65 & Tip of the Red Giant Branch& & Sakai 2000&18.59 $\pm$  0.09$\pm$  0.16 \nl 
66 &  &  &  Romaniello et al 2000  &  18.69 $\pm$  0.25 $\pm$  0.06  \nl 
67 & Trigonometric parallax  &  RR Lyrae & this paper 2001 & 18.53  $\pm$  0.10\tablenotemark{c} \nl 
68 &    &  RR Lyrae & this paper 2001 & 18.38  $\pm$  0.10\tablenotemark{d} \nl 
69 &  & RR Lyrae  &  Luri et al.  1998   &  18.37 $\pm$ 0.23  \nl  
70 &  & RR Lyrae  &  McNamara  1997   &  18.57 $\pm$ 0.03  \nl  
71 &  & Cepheids &  Groenewegen \& Oudmaijer 2000  &  18.45 $\pm$ 0.18  \nl    
71 &  &  &  & to   \nl 
71 &  &   &     &  18.86 $\pm$ 0.12 \nl 
72 &  & Cepheids  &  Groenewegen \& Oudmaijer 2000  &  18.60 $\pm$ 0.11 \tablenotemark{e} \nl  
73 &  & Cepheids  &  Groenewegen \& Oudmaijer 2000  &  18.52 $\pm$ 0.18 \tablenotemark{f} \nl      
74 &  & Cepheids  &  Groenewegen \& Salaris 1999  &  18.61  $\pm$ 0.28 \nl  
75 &  & Cepheids &  Feast 1999  &   18.68 $\pm$ 0.22 \tablenotemark{g} \nl  
76 &  & Cepheids &  Oudmaijer et al. 1998  &    18.56 $\pm$ 0.08 \nl 
77 &  & Cepheids &  Madore \& Freedman  1998 &   18.44 $\pm$ 0.35 \nl  
77 &  &  &  & to   \nl  
77 &  &  &  &  18.57 $\pm$ 0.11 \nl  
78 &  & Cepheids  &  Luri et al.  1998   &  18.29 $\pm$ 0.17  \nl  
79 &  & Cepheids &  Feast \& Catchpole 1997  & 18.70 $\pm$ 0.10 \nl  
80 &  & Cepheids &  Paturel et al. 1997  & 18.72 $\pm$ 0.05 \nl  
81 &  &  HB  &  Carretta  et al. 2000 &18.49 $\pm$ 0.11 \nl
82 &  &  HB  &Gratton 1998 &  18.49 $\pm$  0.11  \nl  
83 &  &  HB &  Koen \& Lacy 1998 &  18.49 $\pm$  0.12   \nl  
84 & White Dwarf cooling sequence   &  & Carretta et al. 2000  &  18.40 $\pm$  0.15 \nl
\enddata        

\tablenotetext{a} { Walker (1992) Photometry }        
\tablenotetext{b} {Udalski (1999) Photometry }        
\tablenotetext{c} {Caretta (2000)  $<${\it V } $>$ = 19.11 (+0.03, $<[$FeH$]>$ 
correction)=19.14}
\tablenotetext{d} {Udalski (1999) $<${\it V } $>$ = 18.94 (+0.05, $<[$FeH$]>$
correction) = 18.99}
\tablenotetext{e} {Based on the {\it PL}-relation in {\it V} \& {\it I} \& the Wesenheit-index}        
\tablenotetext{f} {Based on the {\it PL}-relation in {\it K} }        
\tablenotetext{g} {Data from Koen \& Laney 1998}        

\end{deluxetable}        
\end{center}

%
%

\clearpage

\clearpage

\clearpage
\clearpage

\clearpage

\clearpage

\end{document}